\documentclass[journal]{IEEEtran}
\usepackage{booktabs}
\ifCLASSINFOpdf
\else
\usepackage[dvips]{graphicx}
\fi
\usepackage{url}
\usepackage{array}
\usepackage{multirow}
\usepackage{graphicx}
\usepackage[backref]{hyperref}
\usepackage{epstopdf}
\usepackage{subfigure}
\usepackage{cleveref}
\usepackage{float}
\begin{document}
	
	\title{Speaker Diarization Based on Multi-channel Microphone Array in Small-scale Meeting}
	
	\author{Yuxuan Du, Ruohua Zhou}

	\maketitle
	
	\begin{abstract}
		In the task of speaker diarization, the number of small-scale meetings accounts for a large proportion. When microphone arrays are employed as a recording device, its spatial information is usually ignored by most researchers. In this paper, inspired by the clustering method combining d-vector and microphone array spatial vector, we proposed a diarization method which using multi-channel microphone arrays for a meeting with no more than 4 speakers. We utilize speech enhancement to preprocess the audio from the microphone array. The Steered-Response Power Phase Transform (SRP-PHAT) algorithm are employed to get more accurate speakers, and apply the number of speakers to recluster the speech segments to achieve better performance. Finally, we fuse our system by DOVER-LAP to get the best result. We evaluated our system on the AMI corpus. Compared with the best experimental results so far, our system has achieved largely improvement in the diarization error rate (DER).
	\end{abstract}
	

	\IEEEpeerreviewmaketitle

	\section{Introduction}
	
	\IEEEPARstart s{peaker} diarization is the process of dividing an audio stream into segments that only contain the voice of speakers and marking them with speaker identity. In a word, the task of speaker diarization aims to solve the problem of "who spoken when" \cite{1}. With the latest development of speech processing technology, speaker diarization are used in many artificial intelligence systems and various fields \cite{2}, such as telephone conversation, broadcast news, meetings, clinical records, etc. Modern speaker embedding technologies, for instance, d-vector \cite{3}, c-vector \cite{4} and x-vector \cite{5}, have proved to be able to capture speaker features well. 
	
 Notably, dedicated hardware such as microphone arrays in conferences often allows multiple audio channels, and this information can be used in sound source localization (SSL) techniques, including two-step algorithms based on TDOA, spatial spectral estimation, and steered-response methods based on beamforming. The previous work tried to calculate the inter channel delay characteristics by using acoustic beamforming information to use spatial information. These characteristics were combined with MFCC at the weighted log likelihood level \cite{6}. Recently, \cite{7} used the most advanced deep learning method to acoustic speaker modeling, and proposed a new method to implement speaker diarization by supplementing d-vector with acoustic beamforming when multi-channel microphone arrays are available. 
 
	In this paper, when multi-channel microphone array is available, we put forward a method which to combine Steered-Response Power Phase Transform (SRP - PHAT) algorithm \cite{8} with speech enhancement to determine the number of small conference speaker. Then we employ the number of people we have obtained to do the clustering in different ways. At the same time, we also propose a speaker diarization method using only microphone array and compare it with speaker diarization based on d-vector \cite{9}. We evaluated our method on the publicly available AMI conference corpus \cite{10}. When the number of speakers is determined by our system, the single system and fusion scoring proposed by us show highly competitive performance, and surpass the recent methods in speaker diarization \cite{{11,12}}.

	\section{Microphone Array Diarization}
	\subsection{Acoustic Beamforming}
	
	The core thought of the controllable beamforming method based on maximum output power is to make a beamforming in each direction. GCC algorithm only uses a pair of microphone signals to estimate the time delay, without using the information of other channels. When the reverberation of speech is strong, the localization of the sound source often gets the wrong result. The principle of SRP-PHAT algorithm is to calculate the sum of the generalized cross correlation (GCC-PHAT) functions of all microphones weighted by phase transformation of the received signal at the sound source localization, and to search the whole sound source space to find the point that can make the maximum SRP value as the estimation of the sound source localization. It can be summarized as the following expression: 
	\begin{eqnarray}
		P_{PHAT}(q)=\sum\limits_{ij}^{M(M-1)/2}GCC-PHAT(\tau ij(q))
	\end{eqnarray}
	where M is the number of microphones in the array. $\tau$ij(q) is the theoretical time delay between the received signals at microphone i and microphone j at a given spatial position q. The position estimation of the sound source can be obtained by searching the $\widehat{q}$ that makes q take the maximum value in the whole sound source space, which is the positioning principle of the controllable beamforming method.
	
	In \cite{7}, all steered-response powers computed from a coarser grid is flattened to a vector of the entire space around the microphone array. Although the system performance can be improved by combining with d-vector in this method, the improved performance is still limited. To exploit the information of the spatial vectors, we select a single global maximum from a high-resolution grid. In our experiment, $P^{PHAT}(q)$ of its steered-response powers is calculated with a frame length of 500ms and a frame shift of 500ms, and $P^{PHAT}(q)$ is calculated 256 times in a space of 360° to obtain the most likely direction of arrival.
	
	\subsection{Speech Enhancement}
	
	In the voiceprint field of deep learning, researchers usually utilize single channel speech enhancement methods to improve speech quality. Generally, it can be divided into time-domain based speech enhancement \cite{13} and frequency-domain based speech enhancement \cite{14}. In the AMI dataset, there is a large noise interference, which may have a huge impact on the directional localizing of the sound source. Therefore, this paper adopts the method of merging full-band and sub-band models proposed by \cite{15}, which is called FullSubNet. Based on the complementary advantages of full-band and sub-band, it connects them and combines their respective advantages for joint training. It can not only focus on the global context, but also capture the signal stability, achieving good results.
	
	\subsection{Number of Speakers Based on Multi-channel Microphone Array}
	
\begin{figure}[H]
		\centering
		\includegraphics[width=3.33in, keepaspectratio]{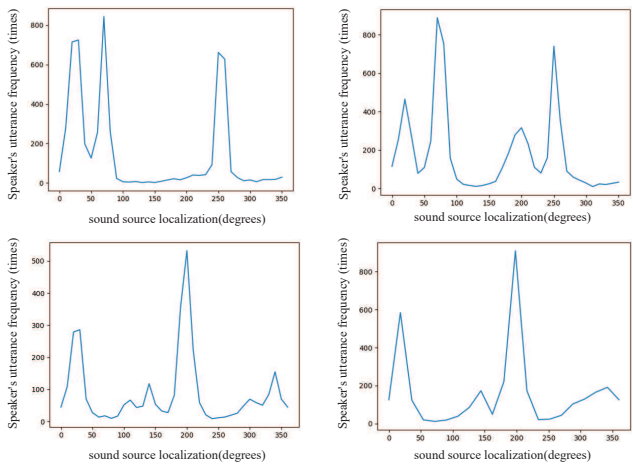}
		\caption{Images regarding speak frequency in AMI meeting. The two images in the first line represent meetings with different numbers of speakers, and the second line represents the number of speakers in a meeting after speech enhancement.}
		\label{fig1}
		
	\end{figure}

	In Section A, we search the largest SRP value point in each sub-segment as the estimation of sound source position. And the number of speaker occurrences is counted every ten degrees in 360 degrees of space. As shown in \cref{fig1}, in the first line we can clearly see the crests corresponding to the three or four local maximum points, which means there are three or four speakers in the meeting. When the data shows a trend from monotonic increase to monotonic decrease, find out the positions of the two maximum points. In order to avoid the dominant position of someone, we specify whether the data of the third or fourth maximum point is greater than one fourth of the second local maximum point (which also means that the new speaker needs to speak more than one fourth of the speaker with the second speaking frequency). The number of local maximum points that satisfy the requirement corresponds to the number of speakers. In the second line, we can see that the number of speakers in the speech enhanced audio has changed from six to four.

Based on this judgment, the accuracy rate of microphone array to determine the number of speakers is shown in \cref{table1}. We calculated the accuracy of the microphone array in predicting the number of speakers before and after speech enhancement. At the same time, we also compare the results of AHC and SC clustering algorithms.
	
	\begin{table}[H]
	\begin{center} 
		\caption{The accuracy of the number of speakers in different clustering methods. Among them, EVAL1 is the AMI dataset of HeadsetMix, and EVAL2 is the AMI dataset of microphone array beamforming. Note that * stands for audio after speech enhancement, and we exclude TNO meetings}
		\label{table1}
		\setlength{\tabcolsep}{3pt}
	
	\begin{tabular}		
		{ccccc}
	 \cmidrule(r){1-5}
		\centering Clustering \par Methods & 
		 \centering Microphone \par Array & 
		 \centering Microphone \par Array* & 
		 AHC & 
		 NME-SC   
		\\ \cmidrule(r){1-5}
		 \centering Eval1 \par Accuracy &
		 83.3$\%$ &
		 100$\%$ & 
		 91.7$\%$ & 
		 83.3$\%$
		\\  \cmidrule(r){1-5}
		 \centering Eval2 \par Accuracy &  83.3$\%$ & 	100$\%$ & 	83.3$\%$ & 75$\%$  \\ \cmidrule(r){1-5}
	
	\end{tabular}
\end{center} 
    \end{table}

	\subsection{Speaker Diarization Based on Multi-channel Microphone Array}
	
\begin{figure}[H]
		\centering
		\includegraphics[width=2.33in, keepaspectratio]{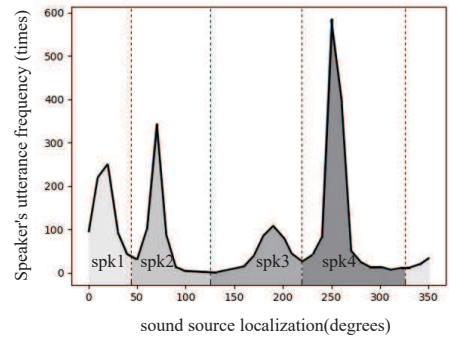}
		\caption{Speaker diarization by finding the timestamp of the speaker in each region.}
		\label{fig2}
	\end{figure}

	A simple speaker diarization method is proposed for multi-channel microphone arrays. First, we use the SRP-PHAT algorithm to obtain the result of the number of speakers. Then, we find out the corresponding angle of each global maximum point and calculate the average angle between two pairs. As shown in \cref{fig2}, we approximately believe that the middle part of the dotted line represents each speaker. We extract all timestamps representing the speaker in the region and finally integrate them into RTTM files.
	It is worth mentioning that the overlap ratio in the AMI dataset is much higher than that in the actual conference, and the noise interference is large. We will get better results when we experiment with clear audio recorded by the microphone array.
	
	\section{Speaker Diarization System}
	
\begin{figure}[H]
	\centering
	\includegraphics[width=3.5in, keepaspectratio]{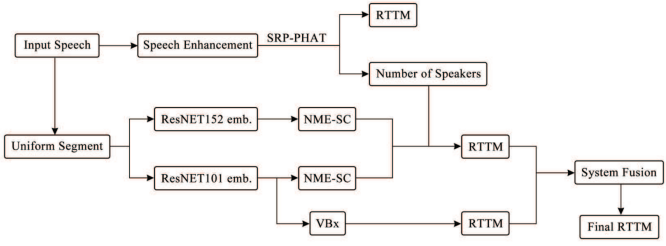}
	\caption{Our proposed speaker diarization pipeline.}
	\label{fig3}
\end{figure}

	Our complete speaker diarization process is shown in \cref{fig3}. Since our research aims at improving the speaker error rate (SER) rather than the performance of voice activity detection (VAD), we firstly use oracle VAD to process the segment of non-speech, and then use SRP-PHAT algorithm to preliminarily determine the number of speakers. We use ResNet 101 and ResNet 152 to extract the speaker embedding. By substituting the number of speakers into the spectrum clustering for re-clustering, the system shows the best single system result. Finally, the spectral clustering and VBx results are fused by DOVER-LAP and scored to obtain better system performance.

	\subsection{Speaker Embedding}
	In speaker diarization systems, embedding words are extracted from homogenous segments of fixed length in utterance \cite{5}. A speaker embedding extractor is employed to transform acoustic features into fixed-dimension feature vectors. In this experiment we utilize ResNet101 and ResNet152 \cite{16} as speaker embedding extractors, whose input is a 64-dimensional log-Mel filter-bank feature with 25ms frame length and 10ms frame shift. Also, the audio was cut into 1.44 second segments with frame shifts of 0.6 and 0.72 seconds. The 16 kHz x-vector extractor is trained using data from VoxCeleb1, VoxCeleb2 \cite{17}, and CN-CELEB. In addition, data from the MUSAN \cite{18} and RIR \cite{19} corpus were additionally augmented.
	
	\subsection{Speaker Clustering}
	
	\subsubsection{Spectral Clustering}
	{We used a novel spectral clustering framework proposed in \cite{20} that can automatically tune the parameters of the clustering algorithm in the context of the speaker diarization. The proposed framework uses the normalized maximum eigengap (NME) value to estimate the number of clusters in the spectral clustering process and the parameters of the element threshold in each row of the affinity matrix without any parameter adjustments to the development set.}
	
	\subsubsection{Agglomerative Hierarchical Clustering}
	{AHC method is often employed in the field of speaker diarization, and there are many ways to measure distance \cite{21}. The method starts with multiple speakers of unknown categories, calculates the similarity scores among each speaker, and merges the similar classes. The above steps are iterated, and the clustering ends when the aggregated categories reach a predetermined threshold.}
	
	\subsubsection{Re-segmentation}
	{The detailed introduction of VBx model is shown in \cite{12}. The preliminary alignment relationship between x-vector and speaker can be obtained through AHC clustering results, which is used to initialize VB parameters, and then the Expectation Maximum (EM) iterative optimization is carried out. In this way, the separation accuracy can be further improved after AHC clustering. This model is used to improve the performance of initial clustering, and the results of initial clustering are used to initialize the VBx model. The hyperparameters mentioned in the VBx model are tuned on the AMI development set, and we choose Fa = 0.3, Fb = 17, loopP = 0.99. }

	\subsubsection{DOVER-Lap based system fusion}
	{Recently, a diarization output voting error reduction (DOVER) method \cite{22} is proposed to combine multiple diarization results based on the voting scheme. In the DOVER method, the speaker labels are aligned one by one between the different diarization systems. After all assumptions are aligned, each system votes its speaker label for each segment region, with each system having a different voting weight, and selects the speaker label that gets the highest voting weight for each segment region. We finally assigned a weight of 0.3, 0.3 and 0.4 to the three RTTM files with the best single system scores.}
	
	\section{Experimental Setup}
	\subsection{Datasets}
	The ResNET101 model is trained on VoxCeleb1 and VoxCelecleb2 data. The RIRs and MUSAN datasets are used for data augmentation. ResNET152 model We use the model provided by the BUT team. For small-scale speaker diarization, we employed the Augmented Multi-party Interaction (AMI) meeting dataset with barely more than four speakers per meeting. The AMI testset including beamformed audio and mix-headset audio. Since the dataset partition criteria in some papers are different, we calculated the diarization error rate of whether to include TNO or not. In addition, the channels in the microphone array are beamformed using the standard BeamformIt toolbox \cite{23}.
	
	\subsection{Diarization Setup}
	Speaker embeddings for each successive speech segment are extracted with a frame length of 1.44 seconds and a frame shift of 0.72 seconds or 0.6 seconds. The clustering method of AHC was used for 0.72 seconds of frame shift, and the threshold was set as -0.015. The 0.6 seconds frame shift is applied in spectral clustering method. The maximum number of speakers estimated by initial clustering is set to 10, and the threshold is automatically adjusted. Finally, after the number of speakers is obtained, the spectral clustering method with known number of clusters is used to re-cluster. In the AMI dataset, we used the standard speaker diarization error rate (DER) as the evaluation index \cite{24}. DER consists of speaker error rate (SER), false alarm (FA), and missed speech (MS), where SER represents errors introduced due to incorrect marking of speaker segments. FA and MS occur due to errors introduced by the Voice Activity Detection (VAD) system. Since this paper focuses on the accuracy of speaker recognition, we use oracle VAD based on ground truth. Similar to \cite{25}, a 0.25 second collar is used and the speaker overlapping area is ignored during the scoring process (also NIST standard). We ultimately scored using a tool developed by NIST for the Rich Transcription 2009 evaluation (NIST RT-09).
	
	\subsection{Results}
	The \cref{table2} summarizes the effect of speaker diarization under beamforming audio streams in the AMI corpus. Since the \cite{7} did not divide the AMI testset according to the official standard, we used the official standard method of dividing the dataset in order to facilitate the follow-up readers to conduct a comparative study of our method. Although there is no consistent dataset partition, it is still referential. In the paper \cite{7}, the author shows the performance of d-vector and its combination with microphone array for speaker diarization. As shown in \cref{table2}, the method that we only use microphone array shows better performance than d-vector under certain comparable conditions. It shows better performance, resulting in 34.7$\%$ and 45.6$\%$ improvement respectively. And after speech enhancement, the error rate of speaker diarization decreases by 2.1$\%$ and 8.0$\%$ respectively. It is worth noting that the method of using only the spatial information of microphone array proposed by us is superior to the method of combining the spatial features of microphone array with d-vector in IS conference. However, this may also be due to the differences in dataset partitioning.
	
	\begin{table}[H]
		\begin{center} 
			\caption{The performance of speaker diarization methods using only microphone arrays}
			\label{table2}
			\setlength{\tabcolsep}{3pt}
			
			\begin{tabular}		
				{cccc}
				\toprule
				\centering \multirow{2}*{Speaker Feature} & 
				\centering \multirow{2}*{Speech Enhancement} & 
				\centering ES Meeting & 
				IS Meeting   
				\\ \cmidrule(r){3-4}
				\centering ~ &
				~ &
				DER/SER & 
				DER/SER 
				\\ \cmidrule(r){1-4}
				\centering Spatial feature &  No & 12.16 & 9.33 \\ \cmidrule(r){1-4}
				\centering Spatial feature &  Yes & 11.91 & $\textbf{8.58}$  \\ \cmidrule(r){1-4}
\centering d-vector only \cite{7} &  - & 18.63 & 17.16  \\ \cmidrule(r){1-4}
\centering Kang et al. \cite{7} &  - & $\textbf{8.76}$ & 11.74  \\ \bottomrule				
			\end{tabular}
		\end{center} 
	\end{table}
	
	In \cref{Table3} we summarizes the effect of speaker diarization under different audio streams in the AMI corpus, where EVAL1 is the HeadsetMix audio stream and EVAL2 is the beamforming audio stream. We have found two papers that have the best results on the AMI dataset so far, but they use different test set partitions. For a fairer comparison, we use two criteria for dividing test sets, that is, whether to exclude TNO meetings ($\#$ in the table means to exclude TNO meetings). Unlike speaker diarization using only microphone arrays, we utilize the completely consistent dataset classification standard (i.e., official standard) for comparison. In addition, we also use standard x-vector embedding for comparison. Since we use the same VAD system for each dataset, the missed detection rate and false alarm rate of different methods are the same. Therefore, speaker confusion is the only criterion to judge the error rate of speaker diarization.
	
	Before using information about the number of speakers, our best experimental results lead to a relative improvement of 31.9$\%$, 69.7$\%$, 27.7$\%$ and 33.2$\%$ respectively compared with the best results in other papers on the four evaluation sets. We use the NME-SC algorithm to re-cluster the number of speakers obtained from microphone array information. The best results in the four evaluation sets lead to a relative improvement of 42.9$\%$, 72.2$\%$, 12.3$\%$ and 50.8$\%$ respectively. Through the fusion algorithm, we compared our best single system results with the fusion results. The performance of the three data sets has been improved, resulting in 8.3$\%$, 7.1$\%$ and 22.8$\%$ relative improvements respectively. These results show that our parameter adjustment and clustering methods can improve the system performance better. And adding the information of the number of speakers will lead to a further decline in the speaker diarization error rate. The best speaker diarization error rate will be obtained by fusing the system output files with the clustering results of a known number of speakers.
	
	\begin{table}
	\begin{center} 
		\caption{The performance of speaker diarization methods using only microphone arrays}
		\label{Table3}
		\setlength{\tabcolsep}{3pt}
		
		\begin{tabular}		
			{ccccccc}
			\toprule
			\centering \multirow{2}*{Embedding} & 
			\multirow{2}*{Cluster} & 
			\multirow{2}*{Re-cluster} &
			Eval-1 &
			Eval-1$\#$ &
			Eval-2 &
			Eval-2$\#$
		
			\\ \cmidrule(r){4-7}
			~ &
			~ &
			~ & 
			DER &
			DER &
			DER &
		    DER
			\\ \cmidrule(r){1-7}
			TDNN &
			NME-SC &
			No &
			— & 
			— & 
			— & 
			11.04
			\\ \cmidrule(r){1-7}
			ECAPA-TDNN &
			NME-SC &
			No &
			— &
			4.03 &
			— &
			3.01
			\\ \cmidrule(r){1-7}
			ResNet101 &
			VBx &
			No &
			2.10 &
			— &
			3.90 &
			—
			\\ \cmidrule(r){1-7}
			ResNet101* &
			VBx &
			No &
			1.50 &
			1.55 &
			3.03 &
			2.38
			\\ \cmidrule(r){1-7}
			ResNet101 &
			NME-SC &
			No &
			1.67 &
			1.51 &
			3.60 &
			2.56
			\\ \cmidrule(r){1-7}
			ResNet152 &
			NME-SC &
			No &
			1.43 &
			1.22 &
			2.82 &
			2.01
			\\ \cmidrule(r){1-7}
			ResNet101 &
			NME-SC &
			Yes &
			1.42 &
			1.26 &
			4.01 &
			1.83
			\\ \cmidrule(r){1-7}
			ResNet152 &
			NME-SC &
			Yes &
			1.2 &
			1.12 &
			3.42 &
			$\textbf{1.48}$
			\\ \cmidrule(r){1-7}
			\multicolumn{3}{c}{Fusion} &
			$\textbf{1.1}$ &
			$\textbf{1.04}$ &
			$\textbf{2.64}$ &
			1.53
			\\
			 \bottomrule				
		\end{tabular}
	\end{center} 
\end{table}
	
	\section{Conclusion}
	In this work, we propose a method to determine the number of small-scale conference speakers using microphone arrays based on Steering Response Power Phase Transform (SRP-PHAT) algorithm. At the same time, we also use our own method to show the performance of speaker diarization using microphone array alone, and the performance is better than the results using d-vector. And after the introduction of speech enhancement, the accuracy of the number of speakers has been improved again. The performance of speaker diarization using the obtained number of speakers exceeds the best results so far, which shows the great potential of using microphone array spatial vectors. In addition, the fusion using the clustering results known to the speaker shows better performance.
	Because square tables are used instead of round tables in the actual scene of the AMI meeting, when the number of speakers increases to five or more, location information becomes a difficult indicator to judge. In future work, we will use microphone arrays to cluster speakers in large conferences to improve its generalization.
	
\section{References}
\bibliographystyle{IEEEtran}
\bibliography{IEEEexample}
	
\end{document}